\documentclass[twocolumn,aps,prl,preprintnumbers,superscriptaddress,reprint,floatfix]{revtex4-1}
\usepackage{bbm}
\usepackage{bm}
\usepackage{amsmath}
\usepackage{amssymb}
\usepackage{empheq}
\usepackage{graphicx}
\usepackage{mathrsfs}
\usepackage{amsfonts}
\usepackage{amsthm}
\usepackage{color}
\usepackage{bigints}
\usepackage{txfonts}
\usepackage{hyperref}
\hypersetup{
     unicode=false,          
     pdftoolbar=true,        
     pdfmenubar=true,        
     pdffitwindow=false,     
     pdfstartview={FitH},    
     pdftitle={My title},    
     pdfauthor={Author},     
     pdfsubject={Subject},   
     pdfcreator={Creator},   
     pdfproducer={Producer}, 
     pdfkeywords={keyword1} {key2} {key3}, 
     pdfnewwindow=true,      
     colorlinks=false,       
     linkcolor=red,          
     citecolor=green,        
     filecolor=magenta,      
     urlcolor=cyan           
}

\providecommand{\U}[1]{\protect\rule{.1in}{.1in}}
\makeatletter
\@ifundefined{textcolor}{}
{
\definecolor{BLACK}{gray}{0}
\definecolor{WHITE}{gray}{1}
\definecolor{RED}{rgb}{1,0,0}
\definecolor{GREEN}{rgb}{0,1,0}
\definecolor{BLUE}{rgb}{0,0,1}
\definecolor{CYAN}{cmyk}{1,0,0,0}
\definecolor{MAGENTA}{cmyk}{0,1,0,0}
\definecolor{YELLOW}{cmyk}{0,0,1,0}
}
\makeatother

\begin{document}
\title{Robustness of Helical Hinge States of Weak Second-Order Topological Insulators}
\author{C. Wang}
\email[Corresponding author: ]{physcwang@tju.edu.cn}
\affiliation{Center for Joint Quantum Studies and Department of 
Physics, School of Science, Tianjin University, Tianjin 300350, China}
\author{X. R. Wang}
\email[Corresponding author: ]{phxwan@ust.hk}
\affiliation{Physics Department, The Hong Kong University of Science 
and Technology (HKUST), Clear Water Bay, Kowloon, Hong Kong}
\affiliation{HKUST Shenzhen Research Institute, Shenzhen 518057, China}
\date{\today}

\begin{abstract}
Robustness of helical hinge states of three-dimensional weak second-order topological 
insulators (WSOTIs) against disorders is studied. The pure WSOTI is obtained from 
a weak $\mathbb{Z}_2$ first-order topological insulator through a surface band inversion. 
Both bulk states and surface states in the WSOTI are gapped, and in-gap valley-momentum 
locked helical hinge states are topologically protected by the surface valley-Chern number. 
In the presence of weak disorders, helical hinge states are robust against disorders while 
the quantized conductance of the states is fragile due to the inter-valley scattering. 
As disorder increases, the system undergoes a series of quantum phase transitions: from the WSOTI to the  
weak first-order topological insulator, then to a diffusive metal and finally to an Anderson insulator.
Our results thus fully establish the WSOTI phase as a genuine state of matters and open a door for the 
second-order valleytronics that allows one to control the valley degree of freedom through helical hinge states.
\end{abstract}

\maketitle

\emph{Introduction.$-$}Topological insulators (TIs) characterized by topological invariants and robust boundary states have 
attracted great interest because of their exotic properties. The band invention resulting in non-zero Chern numbers of a band 
is the central theme of topological materials. The non-zero topological invariants give rise to a bulk-boundary correspondence 
and the necessity of gapless boundary states in the band gap. The standard paradigm of the first-order TIs (FOTIs) claims that a 
$d-$dimensional insulator with band inversion has $(d-1)-$dimensional in-gap boundary states 
\cite{haldane_prl_1988,kane_prl_2005,kane_prl_20051,bernevig_science_2006,konig_science_2007,Roth_science_2009,
hasan_rmp_2010,Moore_nature_2010, qi_rmp_2011,chang_science_2013}. In three-dimensions (3D),  FOTIs are strong (weak) 
when the number of surface Dirac cones is odd (even). A weak FOTI (WFOTI) has zero principle $\mathbb{Z}_2$ index $\nu_0$, 
at least one non-zero weak indexes $(\nu_1,\nu_2,\nu_3)$, and an even number of Dirac cones on surfaces not perpendicular to 
$(\nu_1,\nu_2,\nu_3)$. With this understanding of FOTIs, most recent activities have been focused on higher-order TIs with a 
generalized bulk-boundary correspondence
\cite{zhang_prl_2013,benalcazar_science_2017,peng_prb_2017,langbehn_prl_2017,song_prl_2017,schindler_sciadv_2018,
ezawa_prl_2018,liu_prl_2019,Zhangrx_prl_2019,zhang_prl_2019,lee_prl_2019,varjas_prl_2019,luo_prl_2019,kudo_prl_2019,
chen_pra_2019,Queiroz_prl_2019,li_npj_2019,chen_prl_2020,araki_prb_2019,su_cpb_2019,agarwala_prr_2020,agarwala_arxiv_2020}.
The new paradigm is that, with band inversions on a $d$-dimensional manifold and its sub-manifolds, gapped bands in the manifold 
and its sub-manifolds of dimensions larger than $d-n$ can have gapless states in a boundary sub-manifold of dimension $d-n$. 
For example, a 3D second-order TI has gapless states on the sample hinges inside its bulk and surface band gaps 
\cite{benalcazar_science_2017,langbehn_prl_2017,song_prl_2017,schindler_sciadv_2018,liu_prl_2019,Zhangrx_prl_2019,
Queiroz_prl_2019}. These hinge states have been predicted and observed in real materials, e.g., bismuth crystals 
\cite{schindler_natphys_2018} and magnetic axion insulator Bi$_{2-x}$Sm$_x$Se$_3$ \cite{yue_natphys_2019}.  
\par 

As a well-accepted paradigm, hinge states appear at the intersections of two surfaces of different topological classes when the 
surface Dirac cones of a 3D FOTI are gapped. Hinge states could be either chiral \cite{langbehn_prl_2017} or helical 
\cite{song_prl_2017,schindler_sciadv_2018,Zhangrx_prl_2019,Queiroz_prl_2019}, depending on whether the number 
of surface Dirac cones is odd or even. Robustness of those states against disorders is a fundamental issue because disorders 
exist inevitable in all materials and hinge states must survive in disorders in order to be a genuine state. Chiral hinge states 
can survive in random media due to the absence of backward scattering \cite{wang_arxiv_2020}, while inter-spin/valley 
scatterings are allowed in helical hinge states and may result in the disappearance of these states through Anderson localizations 
at an infinitesimally weak disorder. The occurrence of this scenario, however, contradicts a general belief that the in-gap hinge 
states should persist at finite disorders until the surface state gap closes \cite{sheng_prl_2006,li_prl_2009,trifunovic_prx_2019}. 
Hence, whether disorder-induced backward scatterings can destroy the helical hinge states is not clear and should be examined.
\par

In this letter, we report a weak second-order TI (WSOTI) generated from a WFOTI through band inversion of surface states and 
with mirror symmetry. Different from other reported helical hinge states
\cite{song_prl_2017,Zhangrx_prl_2019,schindler_sciadv_2018,Queiroz_prl_2019} that lock the momentum with spins, carriers 
in different valleys of WSOTIs move to the opposite directions along a hinge. In clean cases, such helical hinge states are characterized by 
the quantized valley-Chern number. They survive in the presence of weak but finite disorders, similar to the surface states in WFOTIs. 
Helical hinge states can be identified by the dominate occupation probability on hinges and negligible occupation probability in the bulks 
and on the surfaces. With increasing disorders, a gap-closing transition from WSOTI to WFOTI happens at a critical disorder $W_{c1}$ at 
which gaps of surface Dirac cones close. Moreover,  with further increasing disorders from $W_{c1}$, the WFOTI becomes a diffusive 
metal (DM), and finally an Anderson insulator (AI). Electronic transport through helical hinge states is also studied. We find that the 
quantum resistance is the sum of an intrinsic contribution from the topological states and an extrinsic part from the inter-valley scatterings 
that is proportional to system sizes. These results cast the authenticity of helical hinge states that provide a way to manipulate the valley 
degree of freedom. 
\par

\begin{figure}[htbp]
\centering
  \includegraphics[width=0.45\textwidth]{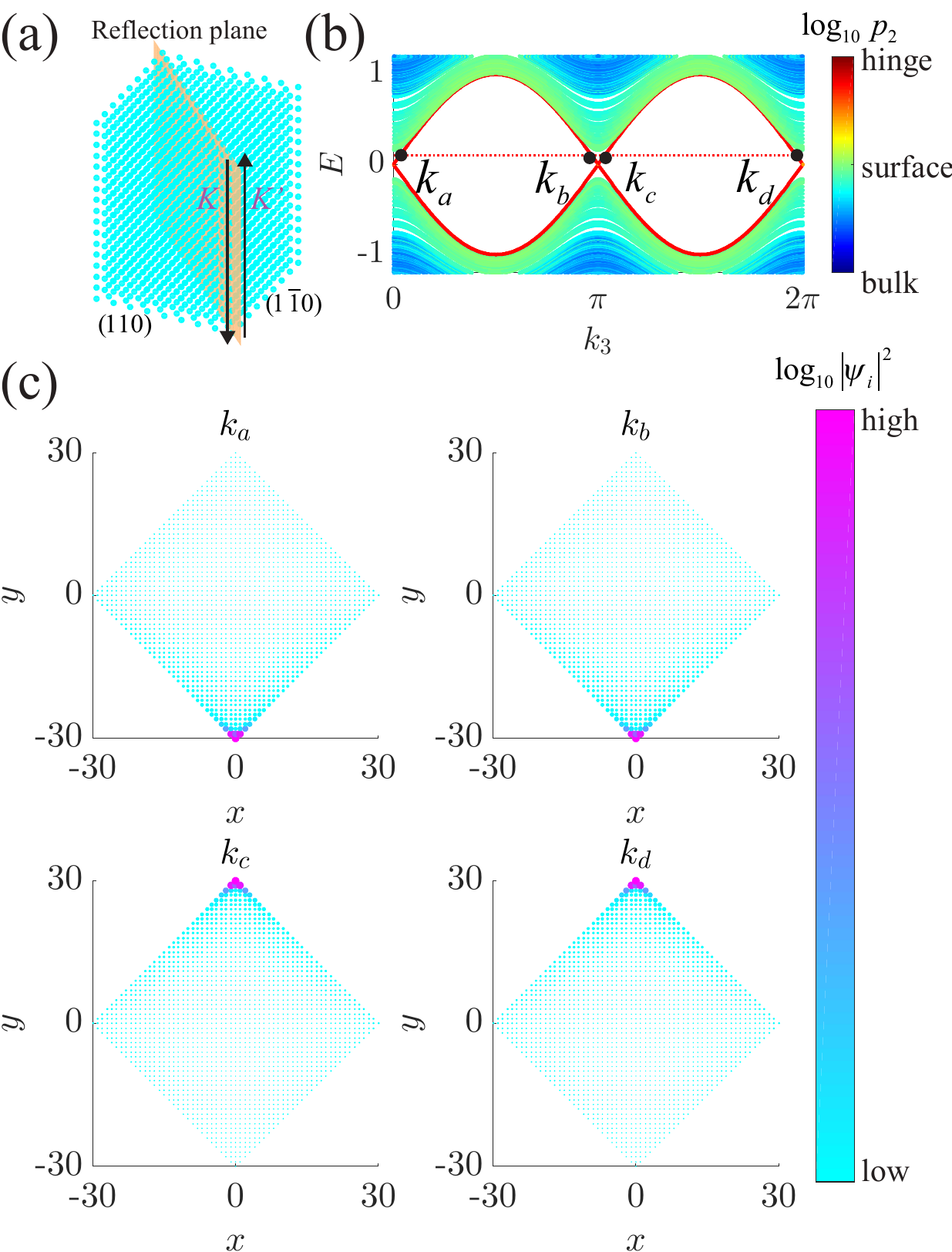}
\caption{(a) Schematic plot of a WSOTI. Hinge states (black arrows) of different valleys are antiparallel. (b) Energy spectrum 
$E(k_3)$ of Eq.~\eqref{hamiltonian2} for $M=t,B=0.2t$. Colors encode $\log_{10}p_2$. Red dotted line locates $E=0.02t$. 
(b) Spatial distribution of the in-gap helical states $k_3=k_{a,b,c,d}$ shown in (b). Colors encode $\log_{10}|\psi_{\bm{i}}|^2$.
}
\label{fig_1}
\end{figure}

\emph{Clean WSOTI.$-$}A clean WSOTI can be modelled by the following Hamiltonian in the momentum space 
\begin{equation}
\begin{gathered}
h_{\text{bulk}}(\bm{k})=t\sin k_2\Gamma^1+(M-t(\cos k_2+\cos k_3))\Gamma^2\\
+t\sin k_3\Gamma^3+t\sin k_1\Gamma^4+B\Gamma^{31}.
\end{gathered}\label{hamiltonian2}
\end{equation}
Here, $\Gamma^{\mu=1,2,3,4,5}=(s_1\otimes\sigma_1,s_2\otimes\sigma_1,s_3\otimes\sigma_1,I_2\otimes\sigma_3,I_2\otimes\sigma_2)$
are the four-by-four non-unique gamma matrices with $s_\mu$ and $\sigma_\mu$ being the Pauli matrices acting on spin and orbital spaces, 
respectively. $I_{\mathcal{D}}$ is the identity matrix of dimension $\mathcal{D}$. Hopping energy $t=1$ is chosen as the energy unit. 
Equation~\eqref{hamiltonian2} is invariant under the reflection symmetry of $\Gamma^{54}=I_2\otimes\sigma_1$, i.e., $\Gamma^{54}
h(k_1,k_2,k_3)\Gamma^{54}=h(-k_1,k_2,k_3)$. For $B=0$ and $M\in(0,2)$, Eq.~\eqref{hamiltonian2} is a reflection-symmetric WFOTI 
with the reflection plane on $x=0$ and characterized by the $\mathbb{Z}_2$ indexes $(\nu_0,\nu_1\nu_2\nu_3)=(0,001)$ 
\cite{fu_prb_2007,chiu_prb_2013,morimoto_prb_2013}. Such $\mathbb{Z}_2$ index guarantees two Dirac cones on the surfaces not 
perpendicular to the vector $(0,0,1)$, e.g., $yz$-facets. For $B>0$, the last term within the surface state subspace acts like an effective 
Dirac mass \cite{langbehn_prl_2017}. To see it, we derive the low-energy effective Hamiltonian of the surface Dirac cones on the 
$yz$-facets with open boundary condition (OBC) applied in the $x-$direction. Such effective Hamiltonian, expanded around 
two Dirac cones (valleys) $\bm{K}=(0,0)$ and $\bm{K}'=(\pm\pi,\pm\pi)$, reads (See Supplemental Materials \cite{supp})
\begin{equation}
\begin{gathered}
h_{\text{surface}}(\bm{p}_\parallel=(p_2,p_3))=
\begin{bmatrix}
h_{\mathbb{S}}(\bm{p}_\parallel) & 0 \\
0 & \mathcal{T}h_{\mathbb{S}}(\bm{p}_\parallel)\mathcal{T}^{-1}
\end{bmatrix}
\end{gathered}\label{hamiltonian3}
\end{equation}
with $h_{\mathbb{S}}(\bm{p}_\parallel)=t p_2\tau_1+tp_3\tau_3+(B- t'^2 p^2_2/2)\tau_2$. $\tau_{1,2,3}$ are the Pauli matrices in the 
basis of the zero-energy surface states wave functions. The upper and lower blocks are for $\bm{K}$ and $\bm{K}'$ related by the pseudo-
time-reversal symmetry represented by $\mathcal{T}=-i\tau_2 \mathcal{C}$ with $\mathcal{C}$ being complex conjugate. In what follows, 
we denote $\bm{K}$ and $\bm{K}'$ by $\eta_i=\pm 1$. If the two facets are separated by a distance, the Newton mass $t'$ decays 
exponentially with the distance such that the band inversion is prevented. While, if they encounter at the reflection plane $x=0$, $t'=t/2$. 
Equation~\eqref{hamiltonian3} has been widely used to describe quantum spin Hall systems, e.g. HgTe/CdTe quantum well, where the 
helical states come in Kramer pairs with spin-momentum locked \cite{bernevig_science_2006}. Analogously, we expect helical states appear 
on the reflection plane as well but with a {\it valley-momentum locked}, i.e., electrons in hinge channels that behave as massless relativistic 
particles with a given valley pseudo-spin is locked to its propagating direction, see Fig.~\ref{fig_1}(a). 
\par

Figure~\ref{fig_1}(b) shows the energy spectrum $E(k_3)$ of model~\eqref{hamiltonian2} on a rectangle 
sample of size $L_\parallel/\sqrt{2}\times L_\parallel/\sqrt{2}\times L_\perp$ with periodic boundary condition 
(PBC) in $z-$direction and OBCs on surfaces perpendicular to $(110)$ and $(1\bar{1}0)$. Colors in 
Fig.~\ref{fig_1}(b) encode the common logarithmic of participation ratio, defined as 
$\mathcal{P}_2(E)=1/\sum_{\bm{i}}|\psi_{\bm{i}}(E)|^{4}$. Here $|\psi_{\bm{i}}(E)|$ is the normalized wave 
function amplitude of energy $E$ at site $\bm{i}$. $\mathcal{P}_2$ measures the number of sites occupied by 
state of $E$ \cite{wang_pra_1989,wang_prl_2015,pixley_prl_2015} and allows one to distinguish hinge states 
from the bulk and surface states easily. Clearly, for $M=t$ and $B=0.2t$, two pairs of gapless hinge modes appear. 
Those near $k_3=0$ ($k_3=\pi$) are described by the up (down) block of Eq.~\eqref{hamiltonian3} \cite{supp}. 
Wave function distributions of four specific hinge states $k_{a,b,c,d}$ of energy $E=0.02t$ are shown in Fig.~\ref{fig_1}(c). 
States of $k_a$ and $k_b$ ($k_c$ and $k_d$), respectively propagating along $\pm z-$directions, are localized on 
the same hinge $x=0,y=L_\parallel/2$ ($x=0,y=-L_\parallel/2$). 
\par

In quantum spin Hall systems where spin $s_z$ is a good quantum number, spin-Chern numbers play the 
role of topological invariant. Similarly, we employ the valley-Chern number $C_{\text{valley}}$ to 
measure the topology of the surface states in clean limit that tells the emergence of helical hinge states. 
$C_{\text{valley}}$, widely used in layered-graphene systems by studying the valley Hall effect 
\cite{zhang_prl_2011,ezawa_prb_2013,zhang_pnas_2013}, is defined as
\begin{equation}
\begin{gathered}
C_{\text{valley}}=C_{\bm{K}}-C_{\bm{K}'}
\end{gathered}\label{topology}
\end{equation}
with $C_{\bm{K}}$ and $C_{\bm{K}'}$ being the valley-Chern number for $\bm{K}$ and
$\bm{K}'$, respectively. The summation of the Berry curvature over all occupied states of
electrons in a valley $\eta_i$ gives $C_{\eta_i}=\eta_i\text{sgn}[B]/2=\pm 1/2$ \cite{supp}.
Thus, the valley-Chern number is quantized to 1.
\par
 
\emph{Stability against disorders.$-$}To study the robustness of the helical hinge states against disorders, we add 
a random on-site potential $V=\sum_{\bm{i}}c^\dagger_{\bm{i}}v_{\bm{i}} I_4 c_{\bm{i}}$ to the lattice model 
of Hamiltonian Eq.~\eqref{hamiltonian2}, where $c^\dagger_{\bm{i}}$ ($c_{\bm{i}}$) is the electron creation 
(annihilation) operator at site $\bm{i}$. $v_{\bm{i}}$ distributes randomly in the range of $[-W/2,W/2]$. Disorders 
break the lattice translational symmetry so that $C_{\text{valley}}$ is not good any more. Yet, we can still use the 
$L_\parallel=L_\perp=L$ dependence of $\eta_{W,L}(E)=\langle\sum_{\bm{i}\in\text{Hinge}}|\psi_{\bm{i},E}(W,L)|^2\rangle$ 
to characterize hinge states, where the sum is over all the lattice sites on two hinges of $x=0,y=\pm L/2$ and $\langle\cdots\rangle$ 
denotes ensemble average. $\eta_{W,L}(E)$ measures the distribution on the hinges. Naturally, for states with dominated
occupation probability on hinges, $\eta_{W,L}(E)$ approaches a finite value for $L\gg 1$; while for surface and bulk states,
$\eta_{W,L}(E)$ should decrease with $L$ algebraically. 
\par 

\begin{figure}[htbp]
\centering
  \includegraphics[width=0.45\textwidth]{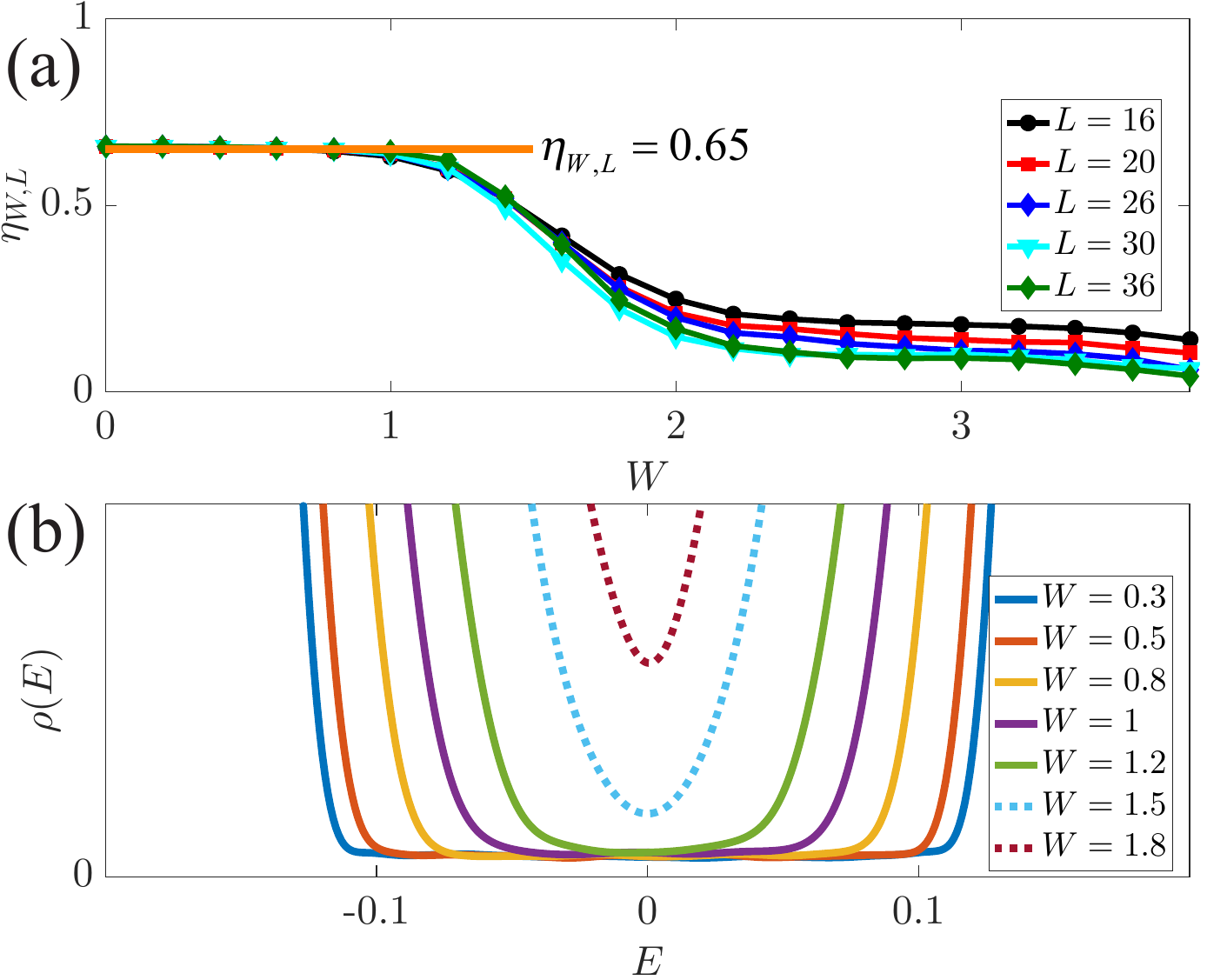}
\caption{(a)  $\eta_{W,L}$ as a function of $W$ for various $L$. 
(b) $\rho(E)$ for $L=200$ and various $W$: solid (dash) lines are for $W<W_{c1}$ ($W>W_{c1}$). 
Here, $M=t$, $B=0.2t$, and $L_\parallel=L_\perp=L$.}
\label{fig_2}
\end{figure}

Let us focus on $E=0$. The obtained $\eta_{W,L}$ as a function of $W$ for various $L$ are shown in Fig.~\ref{fig_2}(a)
\cite{wf,kwant,scipy}. Apparently, there exists a critical disorder $W_{c1}/t\simeq 1.2\pm0.2$ 
below which all curves merge and form a plateau at $\eta_{W,L}\simeq 0.65$, see the orange line. Mergence and 
plateau of $\eta_{W,L}$ are strong indications of helical hinge states at a finite disorder. For $W>W_{c1}$, $\eta_{W,L}$ 
decreases with $L$. As shown below, they are surface states for $W>W_{c1}$, featured by a finite size-independent 
occupation probability on surfaces as $L\to\infty$. 
\par

More insights can be obtained by investigating how disorders affect the gap of surface states through the self-consistent Born 
approximation (SCBA) \cite{chen_prl_2015,liu_prl_2016}, where the self-energy is given by 
\begin{equation}
\begin{gathered}
\Sigma=W^2/(48\pi^2)\int_{\text{BZ}} [(E+i0)I_4-h_{\text{surface}}(\bm{p}_\parallel)-\Sigma]^{-1}d\bm{p}_\parallel.
\end{gathered}\label{self_energy_1}
\end{equation}
We write $\Sigma$ as $\Sigma=\Sigma_0 I_4+\sum^{5}_{\mu=1}\Sigma_\mu\gamma^\mu$ with $\gamma^{1,2,3,4,5}=
(\tau_1\otimes I_2,\tau_3\otimes I_2,\tau_2\otimes\sigma_3,\tau_2\otimes\sigma_3,\tau_1\otimes\sigma_3,\tau_3\otimes\sigma_3)$. 
For $E=0$, $\Sigma_{1,3,4,5}=0$ and $\Sigma_0$ is a pure imaginary number, i.e., $\Sigma_0=i(-1/\tau)$. Then, we obtain 
\cite{supp,shindou_prb_2009}
\begin{equation}
\begin{gathered}
\dfrac{1}{\tau}=\dfrac{1}{\tau}\dfrac{W^2}{48\pi^2t^2}\int_{\text{BZ}}\dfrac{d\bm{p}_\parallel}
{p^2_2+p^2_3+(\tilde{B}-p^2_2/4)^2-1/\tau^2},
\end{gathered}\label{self_energy_2}
\end{equation}
where the Dirac mass is renormalized as $\tilde{B}=B+\Sigma_2$. Here, $\tau$ is the life-time of the zero-energy surface states, i.e., 
$\rho_{\text{surface}}(E=0)\propto 1/\tau$. For $W<W_{c1}$, $1/\tau=0$ and surface states are gapped at $E=0$. While for 
$W>W_{c1}$, finite $\tau$ solutions are allowed and $\rho_{\text{surface}}(E=0)\neq 0$. Thus, with increasing $W$, we expect 
the WSOTI undergoes a gap-closing transition at the critical disorder $W_{c1}$ whose approximate solution is determined from 
Eq.~\eqref{self_energy_2} with $\tilde{B}=B$ is $W_{c1}=t(24\pi/(\ln[16\sqrt{2}/B]))^{1/2}$ \cite{supp}. The closed-form 
solution indicates that $W_{c1}$ increases with $B$, which measures the width of surface gap, and explains qualitatively 
Fig.~\ref{fig_2}(a).
\par

\begin{figure}[htbp]
\centering
  \includegraphics[width=0.45\textwidth]{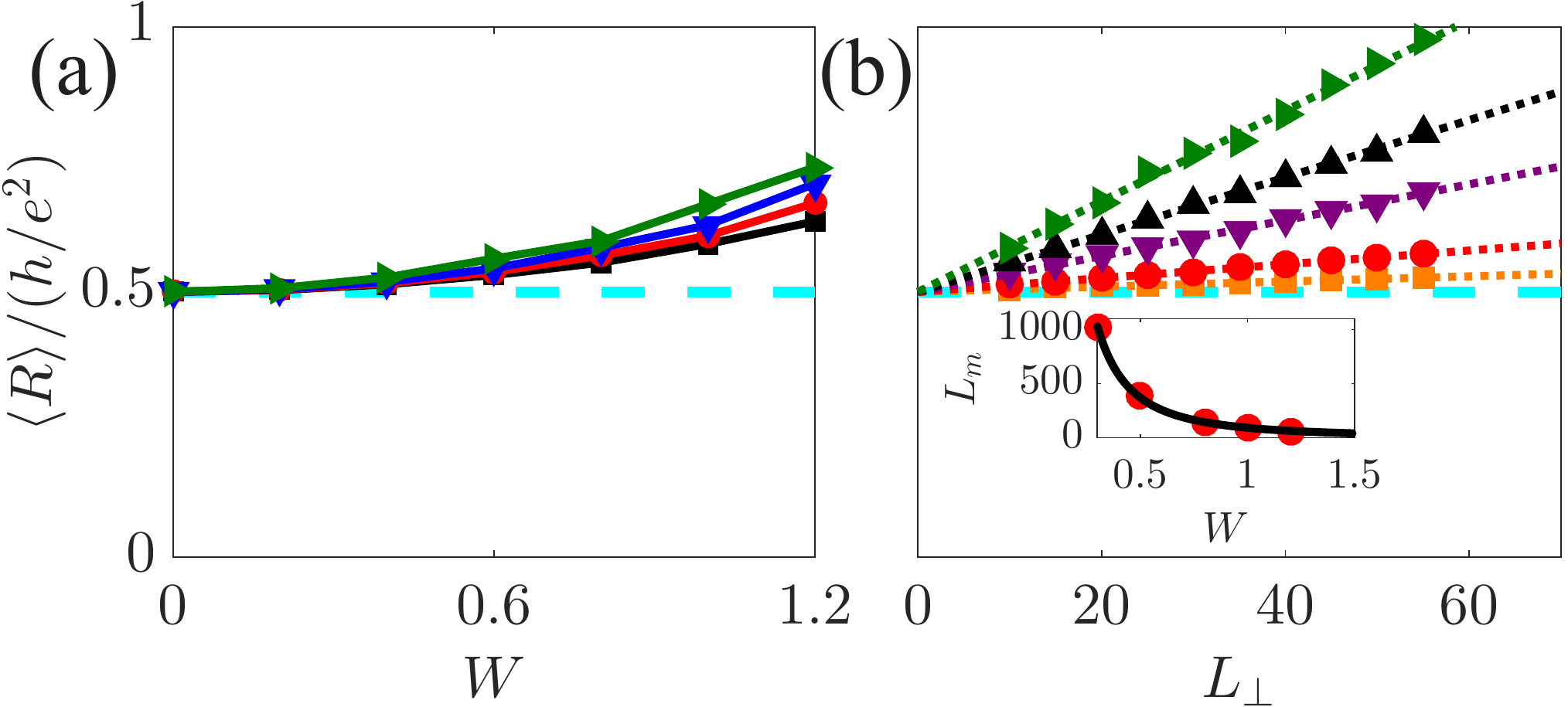}
\caption{(a) $\langle R\rangle$ as a function of $W$ for $L_\perp=L_\parallel=L=16,20,24,32$ (from down to up). (b) 
$\langle R\rangle$ versus $L_\perp$ for $W/t=0.3,0.5,1,0.8,1.2$ (from down to up) and $L_\parallel=10$. Dot lines are 
fitted by Eq.~\eqref{transport1}. Inset: The obtained $L_m$ as a function $W$. Black solid line is a fit of 
$L_m=ct^2/W^2$ with $c=92$. Cyan dash lines locate the intrinsic resistance $h/(2C_{\text{valley}}e^2)$. 
Here, $M=t$ and $B=0.2t$.}
\label{fig_3}
\end{figure}

Dispersion relation of the hinge states in clean limit is linear in $p_3$ near two valleys (see Supplemental Materials \cite{supp}). 
Since disorders do not change the linear dispersion relation within the framework of SCBA, we expect a constant density of 
helical hinge states for $|E|<\Delta_1$ with $\Delta_1$ being the gap of surface states for $W<W_{c1}$. This behavior is 
confirmed by numerical calculations of the average density of states (DOS), defined as 
$\rho(E)=\langle (\sum^{8L^3}_{q=1}\delta(E-E_{q}))\rangle/(8L^3) $ with $E_q$ being the eigenvalues of the systems. We 
calculate $\rho(E)$ through the kernel polynomial method \cite{kpm,DOS} and plot those for $L=200$ and various $W/t$ from 
0.3 to 1.8 in Fig.~\ref{fig_2}(b). Indeed, $\rho(E)$ is independent of $W$ and $E$ within $|E|<\Delta_1$ and $W<W_{c1}$, 
while $\Delta_1$ decreases with $W$. For $W>W_{c1}\simeq 1.2t$, the constant $\rho(E)$ fades, and $\rho(E=0)$ increases 
with $W$. Hence, the constant DOS can be another fingerprint of the helical hinge states, akin to chiral hinge states 
\cite{wang_arxiv_2020}. 
\par

\emph{Electronic transport.$-$}We have also investigated the electronic transport through helical 
hinge states by using the Landauer-Bttiker formula \cite{conductance,macKinnon_zphb_1985} to calculate 
the two-terminal resistance $R$ of the Hall bar connected by two semi-infinite leads along $z-$direction. 
We focus on $W<W_{c1}$ and $E=0$. Figure~\ref{fig_3}(a) plots the 
$\langle R\rangle$ versus $W$ for various $L_\parallel=L_\perp=L$. For $W=0$, $R$ displays perfect quantum 
plateau at $h/(2e^2)$. In the presence of disorders, $\langle R\rangle$ notably increases with $W$ and $L$, 
even for very small disorders. Furthermore, we investigate how $\langle R\rangle$ depends on system sizes. 
Figure~\ref{fig_3}(b) shows $\langle R\rangle$ as a function of $L_\perp$ for various $W$ and a fixed 
$L_\parallel$. We find that $\langle R\rangle$ is linearly increased with $L_\perp$ and can be well 
described by the following formula
\begin{equation}
\begin{gathered}
\langle R\rangle=\dfrac{h}{e^2}\left(\dfrac{1}{2C_{\text{valley}}}+\dfrac{L_\perp}{L_m}\right)
\end{gathered}\label{transport1}
\end{equation}
with $L_m$ being a characteristics length, but independent of $L_\parallel$, see data in 
Supplemental Materials \cite{supp}. Remarkably, very similar features have also been observed 
in quantum spin Hall systems with spin dephasings \cite{jiang_prl_2009}. 
\par 

Equation~\eqref{transport1} can be understood as follows. Unlike chiral hinge states,  
helical hinge states always suffer from the inter-valley scattering caused by short-range 
disorders such that the resistance plateau at $W=0$ are destroyed. Indeed, one can 
treat Eq.~\eqref{transport1} as a combination of an intrinsic resistance $h/(2C_{\text{valley}}e^2)$ 
coming from the non-trivial topology of surface states and  an extrinsic resistance 
due to the inter-valley scattering. The latter should be proportional to $L_\perp$ and independent
of $L_\parallel$. While, $L_m$ is a length acting like mean free length, i.e., $L_m\sim v_g \tau_m$ 
with $1/\tau_m$ being the inter-valley scattering rate and $v_g$ being the group velocity. 
Through Fermi Golden rule, we obtain $L_m\sim t^2/W^2$ (see Supplemental Materials \cite{supp}), which accords 
well with numerical data, as shown in the inset of Fig.~\ref{fig_3}(b).
\par

\begin{figure}[htbp]
\centering
  \includegraphics[width=0.45\textwidth]{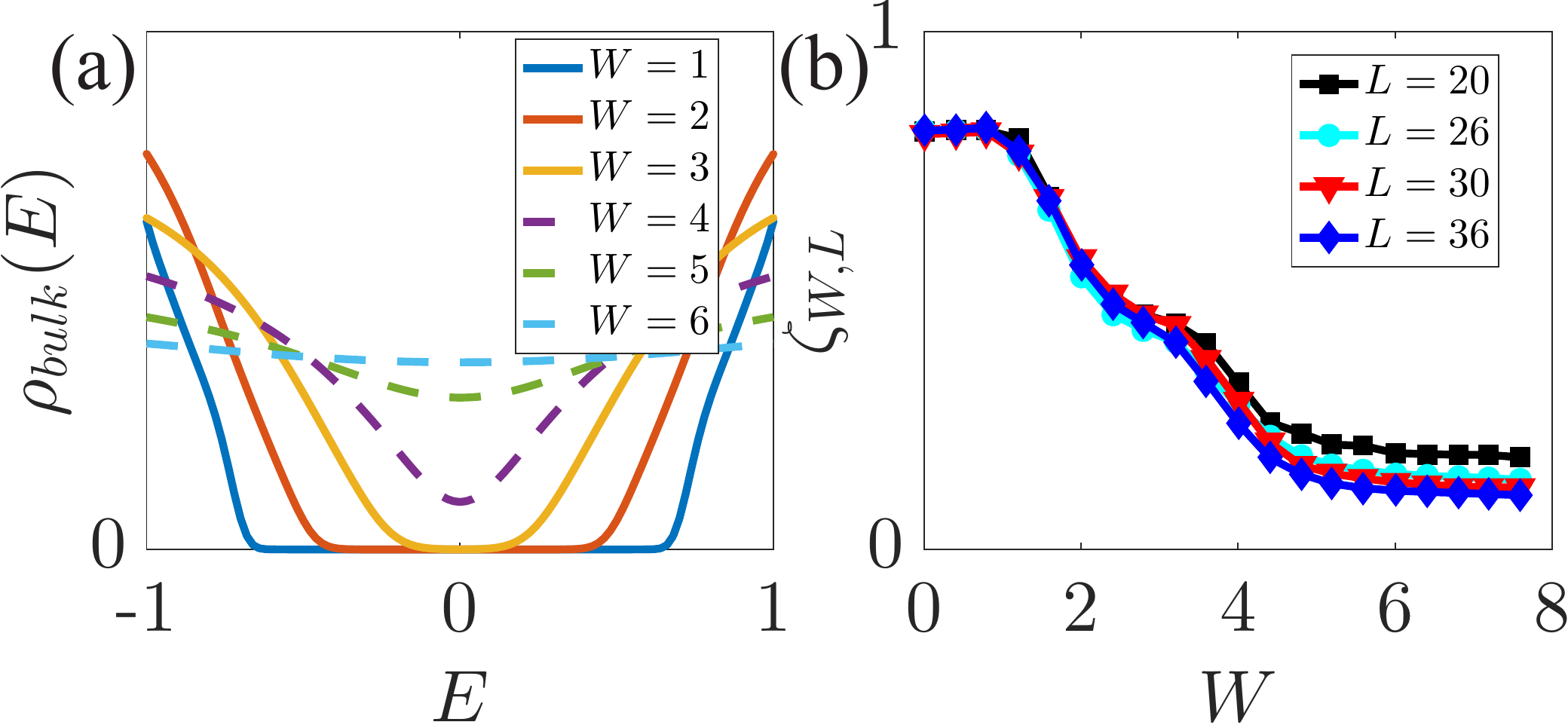}
\caption{(a) $\rho_{\text{bulk}}(E)$ for various $W>W_{c1}$ and $L=L_\parallel=L_\perp=200$. Solid (dash) lines are for $W<W_{c2}$ ($W>W_{c2}$). (b) $\zeta_{W,L}$ versus $W$ for various  $L=L_\parallel=L_\perp$. Here, $M=t$ and $B=0.2t$.}
\label{fig_4}
\end{figure}

\emph{Strong disorders.$-$}To have a complete picture, we study the fate of WSOTIs under stronger disorders. 
For $W>W_{c1}$, the surface energy gap $\Delta_1$ is closed while the bulk energy gap $\Delta_2$ remains finite. 
The system becomes a WFOTI. The conclusion is confirmed by demonstrating that the mid-bulk-gap states are localized 
on the surfaces. Akin to WSOTIs, WFOTIs survive up to a higher disorder $W_{c2}$ at which $\Delta_2=0$ and the system 
transforms into a DM beyond $W_{c2}$. Figure~\ref{fig_4}(a) shows the calculated density of bulk states $\rho_{\text{bulk}}(E)$, 
obtained by applying with PBCs on all directions so that no surface and hinge states are allowed, for various disorders $W>W_{c1}$. 
Clearly, there is always a finite bulk gap for $W<W_{c2}\simeq 3t$. Also, these results demonstrate that the non-zero $\rho(E)$ around 
$E=0$ for $W>W_{c1}$ shown in Fig.~\ref{fig_2}(b) is from the contributions of surface states.
\par

Stronger evidence of the WFOTI-DM transition is given in Fig.~\ref{fig_4}(b), which displays 
$\zeta_{W,L}=\langle\sum_{\bm{i}\in\text{Surface}}|\psi_{\bm{i},E=0}(W,L)|^2\rangle$ as a function of $W$ for various 
$L_\parallel=L_\perp=L$. One should not be confused $\zeta_{W,L}$, the distribution of state $E=0$ on surfaces, with 
$\eta_{W,L}$ of the distribution on hinges. The identification of the nature of state $E=0$ thus can be guided by the 
following observations: (1) For hinge and surface states, $\zeta_{W,L}$ proceeds toward a finite constant in $L\to\infty$; 
(2) For bulk states, $\zeta_{W,L}$ decreases with $L$ and scales with $L$ as $1/L$ for large enough systems. Following 
such criteria, we determine $W_{c2}\simeq 3t$ such that the system is  a WFOTI for $W_{c2}>W\geq W_{c1}$, while 
becomes a DM for $W\geq W_{c2}$. 
\par

Anderson localization occurs at an extremely strong disorders $W_{c3}>W_{c2}$, and the system 
becomes an insulator for $W>W_{c3}$.  We numerically determine $W_{c3}/t=20\pm1$ and the critical 
exponent $\nu=1.5\pm0.1$ through the finite-size scaling analysis of the ensemble-average PR,
$\mathcal{P}_2(E=0)$, see Supplemental Materials \cite{supp}. 
The obtained critical exponent $\nu$ is closed to that of Gaussian unitary ensemble 
established before \cite{wang_arxiv_2020,kawarabayashi_prb_1998}.
\par

\emph{Material relevance.$-$}The WSOTI is a direct consequence of the band inversion of surface 
states of WFOTIs. Remarkably, a recent experiment verified the emergency of  WFOTI phase in 
quasi-one-dimensional bismuth iodide with the same $Z_2$-index studied here \cite{noguchi_nature_2019}. 
Besides, it is found that band inversion of surface states can happen in bismuth with 
respect to certain crystal symmetries \cite{schindler_natphys_2018,yue_natphys_2019}. 
We thus expect bismuth is an ideal material to search for the helical hinge states. 
Rather than electronic systems, WSOTIs may be also found in other systems like photons, 
where the WFOTI has already been visualized \cite{yang_nature_2019} and a band inversion can 
be artificially induced in principle.
\par

\emph{Conclusion.$-$}In short, we have theoretically demonstrated the genuineness of WSOTIs with 
valley-momentum locked helical hinge states. Such hinge states are featured by the quantized 
valley-Chern number in clean limit and are robust against disorders until the band gap of surface 
states collapses. However, the normal quantized conductance of 1D channel is destroyed  by disorders. 
With further increasing disorder, quantum transitions from WSOTI to WFOTI and from WFOTI to DM 
happen in order. At very strong disorders, the system becomes a insulator through the Anderson localization transition.
\par

\begin{acknowledgments}
This work is supported by the National Natural Science Foundation of China (Grants No.~11774296, 11704061 and 11974296) 
and Hong Kong RGC (Grants No.~16301518 and 16301619). CW acknowledges the kindly help from Jie Lu.
\par

\end{acknowledgments}


\begin{thebibliography}{99}


\bibitem{haldane_prl_1988}
F. D. M. Haldane, Model for a Quantum Hall Effect without Landau Levels: Condensed-Matter Realization 
of the ``Parity Anomaly'', 
\href{https://journals.aps.org/prl/abstract/10.1103/PhysRevLett.61.2015}
{Phys. Rev. Lett. {\bf 61}, 2015 (1988)}.


\bibitem{kane_prl_2005} 
C. L. Kane and E. J. Mele, ${Z}_{2}$ Topological Order and the Quantum Spin Hall Effect,
\href{https://journals.aps.org/prl/abstract/10.1103/PhysRevLett.95.146802}
{Phys. Rev. Lett. {\bf 95}, 146802 (2005)}.


\bibitem{kane_prl_20051} 
C. L. Kane and E. J. Mele, Quantum Spin Hall Effect in Graphene,
\href{https://journals.aps.org/prl/abstract/10.1103/PhysRevLett.95.226801}
{Phys. Rev. Lett. {\bf 95}, 226801 (2005)}.


\bibitem{bernevig_science_2006}
B. A. Bernevig, T. L. Hughes, and S.-C. Zhang, Quantum Spin Hall Effect and Topological Phase Transition in 
HgTe Quantum Wells, 
\href{https://science.sciencemag.org/content/314/5806/1757}
{Science {\bf 314}, 1757 (2006)}.

\bibitem{konig_science_2007}
M. K\"{o}nig, S. Wiedmann, C. Br\"{o}une, A. Roth, H. Buhmann, L. W. Molenkamp, X.-L. Qi, and S.-C. Zhang, Quantum Spin Hall Insulator State in HgTe Quantum Wells, 
\href{https://science.sciencemag.org/content/318/5851/766}
{Science {\bf 318}, 766 (2007)}.

\bibitem{Roth_science_2009}
A. Roth, C. Br\"{u}ne, H. Buhmann, L. W. Molenkamp, J. Maciejko, X.-L. Qi, S.-C. Zhang, 
Nonlocal Transport in the Quantum Spin Hall State, 
\href{https://science.sciencemag.org/content/325/5938/294.abstract}
{Science {\bf 325}, 294 (2009)}.

\bibitem{hasan_rmp_2010}
M. Z. Hasan and C. L. Kane, Colloquium: Topological insulators,
\href{https://journals.aps.org/rmp/abstract/10.1103/RevModPhys.82.3045}
{Rev. Mod. Phys. {\bf 82}, 3045 (2010)}.

\bibitem{Moore_nature_2010}
J. E. Moore, The birth of topological insulators,
\href{https://www.nature.com/articles/nature08916}
{Nature (London) {\bf 464}, 194 (2010).}

\bibitem{qi_rmp_2011}
X.-L. Qi and S.-C. Zhang, Topological insulators and superconductors,
\href{https://journals.aps.org/rmp/abstract/10.1103/RevModPhys.83.1057}
{Rev. Mod. Phys. {\bf 83}, 1057 (2011)}.

\bibitem{chang_science_2013}
C.-Z. Chang, J. Zhang, X. Feng, J. Shen, Z. Zhang, M. Guo, K. Li, Y. Ou, P. Wei, L.-L. Wang, Z.-Q. Ji, 
Y. Feng, S. Ji, X. Chen, J. Jia, X. Dai, Z. Fang, S.-C. Zhang, K. He, Y. Wang1, L. Lu, X.-C. Ma, 
Q.-K. Xue, Experimental Observation of the Quantum Anomalous Hall Effect in a Magnetic Topological Insulator,
\href{https://science.sciencemag.org/content/340/6129/167}
{Science {\bf 340}, 167 (2013)}.


\bibitem{zhang_prl_2013}
F. Zhang, C.L. Kane, and E.J. Mele, Surface State Magnetization and Chiral Edge States on 
Topological Insulators,
\href{https://journals.aps.org/prl/abstract/10.1103/PhysRevLett.110.046404}
{Phys. Rev. Lett. {\bf 110}, 046404 (2013)}.

\bibitem{benalcazar_science_2017}
W. A. Benalcazar, B. A. Bernevig, and T. L. Hughes, 
Quantized electric multipole insulators, 
\href{https://science.sciencemag.org/content/357/6346/61.abstract}
{Science {\bf 357}, 61 (2017)}.

\bibitem{peng_prb_2017}
Y. Peng, Y. Bao, and F. von Oppen, Boundary Green functions of topological insulators and superconductors,
\href{https://journals.aps.org/prb/abstract/10.1103/PhysRevB.95.235143} 
{Phys. Rev. B {\bf 95}, 235143 (2017)}.

\bibitem{langbehn_prl_2017}
J. Langbehn, Y. Peng, L. Trifunovic, F. von Oppen, and P. W. Brouwer, Reflection-Symmetric Second-Order 
Topological Insulators and Superconductors, 
\href{https://journals.aps.org/prl/abstract/10.1103/PhysRevLett.119.246401} 
{Phys. Rev. Lett. {\bf 119}, 246401 (2017)}.

\bibitem{song_prl_2017}
Z. Song, Z. Fang, and C. Fang, $(d-2)$-Dimensional Edge States of Rotation Symmetry Protected Topological States, 
\href{https://journals.aps.org/prl/abstract/10.1103/PhysRevLett.119.246402}
{Phys. Rev. Lett. {\bf 119}, 246402 (2017)}.

\bibitem{schindler_sciadv_2018}
F. Schindler, A. M. Cook, M. G. Vergniory, Z. Wang, S. S. P. Parkin, B. A. Bernevig, and T. Neupert, 
Higher-Order Topological Insulators, 
\href{https://advances.sciencemag.org/content/4/6/eaat0346}
{Sci. Adv. {\bf 4}, eaat0346 (2018)}.

\bibitem{ezawa_prl_2018}
M. Ezawa, 
Higher-Order Topological Insulators and Semimetals
on the Breathing Kagome and Pyrochlore Lattices, 
\href{https://journals.aps.org/prl/abstract/10.1103/PhysRevLett.120.026801}
{Phys. Rev. Lett. {\bf 120}, 026801 (2018)}.

\bibitem{liu_prl_2019}
T. Liu, Y.-R. Zhang, Q. Ai, Z. Gong, K. Kawabata, M. Ueda, and F. Nori, 
Second-Order Topological Phases in Non-Hermitian Systems,
\href{https://journals.aps.org/prl/abstract/10.1103/PhysRevLett.122.076801} 
{Phys. Rev. Lett. {\bf 122}, 076801 (2019)}.

\bibitem{Zhangrx_prl_2019}
R.-X. Zhang, W. S. Cole, and S. Das Sarma,
Helical Hinge Majorana Modes in Iron-Based Superconductors,
\href{https://journals.aps.org/prl/abstract/10.1103/PhysRevLett.122.187001}
{Phys. Rev. Lett. {\bf 122}, 187001 (2019)}.

\bibitem{zhang_prl_2019}
Z. Zhang, M. R. Lopez, Y. Cheng, X. Liu, and J. Christensen,
Non-Hermitian Sonic Second-Order Topological Insulator,
\href{https://journals.aps.org/prl/abstract/10.1103/PhysRevLett.122.195501}
{Phys. Rev. Lett. {\bf 122}, 195501 (2019)}.

\bibitem{lee_prl_2019}
C. H. Lee, L. Li, and J. Gong, Hybrid Higher-Order Skin-Topological Modes in Nonreciprocal Systems,
\href{https://journals.aps.org/prl/abstract/10.1103/PhysRevLett.123.016805}
{Phys. Rev. Lett. {\bf 123}, 016805 (2019)}.

\bibitem{Queiroz_prl_2019}
R. Queiroz and A. Stern, Splitting the Hinge Mode of Higher-Order Topological Insulators,
\href{https://journals.aps.org/prl/abstract/10.1103/PhysRevLett.123.036802}
{Phys. Rev. Lett. {\bf 123}, 036802 (2019)}.

\bibitem{luo_prl_2019}
X.-W. Luo and C. Zhang, Higher-Order Topological Corner States Induced by Gain and Loss,
\href{https://journals.aps.org/prl/abstract/10.1103/PhysRevLett.123.073601} 
{Phys. Rev. Lett. {\bf 123}, 073601 (2019)}.

\bibitem{varjas_prl_2019}
D. Varjas, A. Lau, K. Pöyhönen, A. R. Akhmerov, D. I. Pikulin, and I. C. Fulga,
Topological Phases without Crystalline Counterparts,
\href{https://journals.aps.org/prl/abstract/10.1103/PhysRevLett.123.196401}
{Phys. Rev. Lett. {\bf 123}, 196401 (2019)}.

\bibitem{kudo_prl_2019}
K. Kudo, T. Yoshida, and Y. Hatsugai, Higher-Order Topological Mott Insulators,
\href{https://journals.aps.org/prl/abstract/10.1103/PhysRevLett.123.196402} 
{Phys. Rev. Lett. {\bf 123}, 196402 (2019)}.

\bibitem{chen_pra_2019}
H. Chen and X. C. Xie, Interaction-driven topological switch in
a $P-$band honeycomb lattice, 
\href{https://journals.aps.org/pra/abstract/10.1103/PhysRevA.100.013601}
{Phys. Rev. A {\bf 100}, 013601 (2019)}.

\bibitem{araki_prb_2019}
H. Araki, T. Mizoguchi, and Y. Hatsugai,
Phase diagram of a disordered higher-order topological insulator: 
A machine learning study, 
\href{https://journals.aps.org/prb/abstract/10.1103/PhysRevB.99.085406}
{Phys. Rev. B 99, 085406 (2019)}.

\bibitem{li_npj_2019}
Z.-X. Li, Y. Cao, P. Yan, X. R. Wang, Higher-order topological solitonic insulators,
\href{https://www.nature.com/articles/s41524-019-0246-4} 
{npj Comput. Mater. {\bf 5}, 107 (2019)}.

\bibitem{su_cpb_2019}
Z. Su, Y. Kang, B. Zhang, Z. Zhang, and H. Jiang, 
Disorder induced phase transition in magnetic higher-order 
topological insulator: A machine learning study,
\href{http://cpb.iphy.ac.cn/article/2019/2012/cpb_28_11_117301.html}
{Chin. Phys. B {\bf 28}, 117301 (2019).}

\bibitem{chen_prl_2020}
R. Chen, C.-Z. Chen, J.-H. Gao, B. Zhou, and D.-H. Xu, Higher-Order Topological Insulators in Quasicrystals,
\href{https://journals.aps.org/prl/abstract/10.1103/PhysRevLett.124.036803}
{Phys. Rev. Lett. {\bf 124}, 036803 (2020)}.

\bibitem{agarwala_prr_2020}
A. Agarwala, V. Juričić, and B. Roy,
Higher-order topological insulators in amorphous solids,
\href{https://journals.aps.org/prresearch/abstract/10.1103/PhysRevResearch.2.012067}
{Phys. Rev. Research {\bf 2}, 012067(R) (2020)}.

\bibitem{agarwala_arxiv_2020}
A. Agarwala and B. Roy,
Higher-order topological insulators in amorphous solids,
\href{https://arxiv.org/abs/2002.09475}
{arXiv:2002.09475}.

\bibitem{schindler_natphys_2018}
F. Schindler, Z. Wang, M. G. Vergniory, A. M. Cook, A. Murani, S. Sengupta, A. Y. Kasumov, R. Deblock, 
S. Jeon, I. Drozdov, H. Bouchiat, S. Gu\'{e}ron, A. Yazdani, B. A. Bernevig, and T. Neupert, 
Higher-order topology in bismuth,
\href{https://www.nature.com/articles/s41567-018-0224-7}
{Nat. Phys. {\bf 14}, 918 (2018)}

\bibitem{yue_natphys_2019}
C. Yue, Y. Xu, Z. Song, H Weng, Y.-M. Lu, C. Fang, and X. Dai,
Symmetry-enforced chiral hinge states and surface quantum anomalous Hall effect in the magnetic axion 
insulator Bi$_{2–x}$Sm$_x$Se$_3$,
\href{https://www.nature.com/articles/s41567-019-0457-0}
{Nat. Phys. {\bf 15}, 577 (2019)}.

\bibitem{wang_arxiv_2020}
C. Wang and X. R. Wang,
\href{https://arxiv.org/abs/2005.06740}
{arXiv:2005.06740v2}.

\bibitem{sheng_prl_2006}
D. N. Sheng, Z. Y. Weng, L. Sheng, and F. D. M. Haldane,
Quantum Spin-Hall Effect and Topologically Invariant Chern Numbers
\href{https://journals.aps.org/prl/abstract/10.1103/PhysRevLett.97.036808}
{Phys. Rev. Lett. {\bf 97}, 036808 (2006)}.

\bibitem{li_prl_2009}
J. Li, R.-L. Chu, J. K. Jain, and S.-Q. Shen,
Topological Anderson Insulator,
\href{https://journals.aps.org/prl/abstract/10.1103/PhysRevLett.102.136806}
{Phys. Rev. Lett. {\bf 102}, 136806 (2009)}.

\bibitem{trifunovic_prx_2019}
L. Trifunovic and P. W. Brouwer, Higher-Order Bulk-Boundary
Correspondence for Topological Crystalline Phases, 
\href{https://journals.aps.org/prx/abstract/10.1103/PhysRevX.9.011012}
{Phys. Rev. X {\bf 9}, 011012 (2019)}.

\bibitem{fu_prb_2007}
L. Fu and C. L. Kane,
Topological insulators with inversion symmetry,
\href{https://journals.aps.org/prb/abstract/10.1103/PhysRevB.76.045302}
{Phys. Rev. B {\bf 76}, 045302 (2007)}.

\bibitem{chiu_prb_2013}
C.-K. Chiu, H. Yao, and S. Ryu,
Classification of topological insulators and superconductors 
in the presence of reflection symmetry,
\href{https://journals.aps.org/prb/abstract/10.1103/PhysRevB.88.075142}
{Phys. Rev. B {\bf 88}, 075142 (2013)}.

\bibitem{morimoto_prb_2013}
T. Morimoto and A. Furusaki,
Topological classification with additional symmetries from Clifford algebras,
\href{https://journals.aps.org/prb/abstract/10.1103/PhysRevB.88.125129}
{Phys. Rev. B {\bf 88}, 125129 (2013)}.

\bibitem{supp}
See Supplemental Materials at http://link.aps.org/supplemental.

\bibitem{wang_pra_1989}
X. R. Wang, Y. Shapir, and M. Rubinstein, 
Analysis of multiscaling structure in diffusion-limited 
aggregation: A kinetic renormalization-group approach,
\href{https://journals.aps.org/pra/abstract/10.1103/PhysRevA.39.5974} 
{Phys. Rev. A {\bf 39}, 5974 (1989)}.

\bibitem{wang_prl_2015}
C. Wang, Y. Su, Y. Avishai, Y. Meir, and X. R. Wang, Band of
Critical States in Anderson Localization in a Strong Magnetic
Field with Random Spin-Orbit Scattering,
\href{https://journals.aps.org/prl/abstract/10.1103/PhysRevLett.114.096803} 
{Phys. Rev. Lett. {\bf 114}, 096803 (2015)}.

\bibitem{pixley_prl_2015}
J.H. Pixley, P. Goswami, and S. Das Sarma,
Anderson Localization and the Quantum Phase 
Diagram of Three Dimensional Disordered Dirac Semimetals,
\href{https://journals.aps.org/prl/abstract/10.1103/PhysRevLett.115.076601}
{Phys. Rev. Lett. {\bf 115}, 076601 (2015)}.

\bibitem{zhang_prl_2011}
F. Zhang, J. Jung, G. A. Fiete, Q. Niu, and A. H. MacDonald,
Spontaneous Quantum Hall States in Chirally Stacked Few-Layer Graphene Systems,
\href{https://journals.aps.org/prl/abstract/10.1103/PhysRevLett.106.156801}
{Phys. Rev. Lett. {\bf 106}, 156801 (2011)}

\bibitem{ezawa_prb_2013}
M. Ezawa, Topological Kirchhoff law and bulk-edge correspondence for valley Chern
and spin-valley Chern numbers,
\href{https://link.aps.org/doi/10.1103/PhysRevB.88.161406}
{Phys. Rev. B {\bf 88}, 161406 (2013)}.

\bibitem{zhang_pnas_2013}
F. Zhang, A. H. MacDonaldb, and E. J. Melea, 
Valley Chern numbers and boundary modes in gapped bilayer graphene,
\href{https://www.pnas.org/content/110/26/10546}
{Proc. Natl. Acad. Sci. (USA) {\bf 110}, 10546 (2013)}.



\bibitem{wf}
We use the retarded Lanczos method to find the eigenfunction 
of the nearest level around $E=0$ of the disordered bar and 
calculate $\eta_{W,L}$ and $\zeta_{W,L}$ accordingly. In our
scenario, we first use the KWANT package \cite{kwant} to construct a 
Hamiltonian matrix $H$ out of tight-binding model Eq.~\eqref{hamiltonian2}. 
We then solve the eigenequation $H\psi=E\psi$ using the 
SCIPY library \cite{scipy} to obtain the required eigenenergies and 
eigenfunctions.

\bibitem{kwant}
C. W. Groth, M. Wimmer, A. R. Akhmerov, and X. Waintal,
Kwant: A software package for quantum transport, 
\href{https://doi.org/10.1088/1367-2630/16/6/063065}
{New J. Phys. {\bf 16}, 063065 (2014)}.

\bibitem{scipy}
P. Virtanen, R. Gommers, T. E. Oliphant, M. Haberland, 
T. Reddy, D. Cournapeau, E. Burovski, P. Peterson, W. Weckesser,
J. Bright {\it et al.}, 
SciPy 1.0: fundamental algorithms for scientific computing in Python,
\href{https://www.nature.com/articles/s41592-019-0686-2} 
{Nat. Methods {\bf 17}, 261 (2020)}.


\bibitem{chen_prl_2015}
C.-Z. Chen, J. Song, H. Jiang, Q.-F. Sun, Z. Wang, and X. C. Xie
Disorder and Metal-Insulator Transitions in Weyl Semimetals,
\href{https://journals.aps.org/prl/abstract/10.1103/PhysRevLett.115.246603}
{Phys. Rev. Lett. {\bf 115}, 246603 (2015)}.

\bibitem{liu_prl_2016}
S. Liu, T. Ohtsuki, and R. Shindou,
Effect of Disorder in a Three-Dimensional Layered Chern Insulator,
\href{https://journals.aps.org/prl/abstract/10.1103/PhysRevLett.116.066401}
{Phys. Rev. Lett. {\bf 116}, 066401 (2016)}.


\bibitem{shindou_prb_2009}
R. Shindou and S. Murakami,
Effects of disorder in three-dimensional $Z_2$ quantum spin Hall systems,
\href{https://journals.aps.org/prb/abstract/10.1103/PhysRevB.79.045321}
{Phys. Rev. B {\bf 79}, 045321 (2009)}.

\bibitem{kpm}
A. Wei\ss{}e, G. Wellein, A. Alvermann, and H. Fehske,
The kernel polynomial method,
\href{https://journals.aps.org/rmp/abstract/10.1103/RevModPhys.78.275}
{Rev. Mod. Phys. {\bf 78}, 275 (2006)}.

\bibitem{DOS}
The average $\rho(E)$ of the disordered bar of $L=200$ for various $W$ 
that are obtained from 10 ensembles with 1024 Chebyshev moments.


\bibitem{conductance}
The dimensionless resistance of a disordered bar between two clean semi-infinite leads at 
a given Fermi level $E$ can be calculated by $R=h/(e^2\text{Tr}[TT^\dagger])$ with $T$ being 
the transmission matrix \cite{macKinnon_zphb_1985}.

\bibitem{macKinnon_zphb_1985}
A. MacKinnon, 
The calculation of transport properties and density of states 
of disordered solids, 
\href{https://link.springer.com/article/10.1007%2FBF01328846}
{Z. Phys. B {\bf 59}, 385 (1985)}.

\bibitem{jiang_prl_2009}
H. Jiang, S. Cheng, Q.-f. Sun, and X. C. Xie,
Topological Insulator: A New Quantized Spin Hall Resistance Robust to Dephasing,
\href{https://journals.aps.org/prl/abstract/10.1103/PhysRevLett.103.036803}
{Phys. Rev. Lett. {\bf 103}, 036803 (2009)}.

\bibitem{kawarabayashi_prb_1998}
T. Kawarabayashi, B. Kramer, and T. Ohtsuki, Anderson transitions in three-dimensional disordered 
systems with randomly varying magnetic flux,
\href{https://journals.aps.org/prb/abstract/10.1103/PhysRevB.57.11842}
{Phys. Rev. B {\bf 57}, 11842 (1998)}.


\bibitem{noguchi_nature_2019}
R. Noguchi, T. Takahashi, K. Kuroda1, M. Ochi, T. Shirasawa, M. Sakano, C. Bareille, M. Nakayama, 
M. D. Watson, K. Yaji, {\it et. al.}, A weak topological insulator state in quasi-onedimensional
bismuth iodide, 
\href{https://doi.org/10.1038/s41586-019-0927-7}
{Nature {\bf 566}, 518 (2019)}.

\bibitem{yang_nature_2019}
Y. Yang, Z. Gao, H. Xue, L. Zhang, M. He, Z. Yang, R. Singh, Y. Chong, B. Zhang, and 
H. Chen, Realization of a three-dimensional photonic topological insulator, 
\href{https://www.nature.com/articles/s41586-018-0829-0}
{Nature {\bf 565}, 622 (2019)}.

\end{thebibliography}
\end{document}